\documentclass[12pt]{iopart}
\usepackage{iopams}
\usepackage{graphicx}  
\usepackage{bm}

\newcommand\ie {{\it i.e. }}
\newcommand\eg {{\it e.g. }}

\newcommand\mbf {\mathbf}

\newcommand\ket [1] {|#1 \rangle }

\newcommand{\bracket}[2]   {  \left<#1 |  #2\right>}
\newcommand{\av}[1]{\langle #1\rangle}

\newcommand{\be}[1]{ \begin{eqnarray} \mbox{$\label{#1}$} }
\newcommand{\ee}{\end{eqnarray}}
\newcommand{\pref}[1]{(\ref{#1})}

\def\a{\alpha}
\def\b{\beta}

\def\zb{\bar z}
\def\wb{\bar w}
\def\xib{\bar \xi}
\def\half{\frac12}

\newcommand{\bvphi}{\bm{\varphi}}

\begin{document}

\title {A general approach to quantum Hall hierarchies }

\author{J. Suorsa$^1$, S. Viefers$^1$, T.H. Hansson$^2$}

\address{$^1$ Department of Physics, University of Oslo, Box 1048 Blindern, NO-0316 Oslo, Norway}

\address{$^2$ Department of Physics, Stockholm University, AlbaNova University Center, SE-106 91 Stockholm, Sweden}

\ead{j.m.suorsa@fys.uio.no}

\begin{abstract}

The abelian hierarchy of quantum Hall states accounts for most of the states in the lowest Landau level, and there is evidence of a similar hierarchy of non-abelian states emanating from the $\nu = 5/2$  Moore-Read state in the second Landau level. Extending a recently developed  formalism for hierarchical quasihole condensation, we present a theory that allows for the explicit construction of the ground state wave function, as well as its quasiparticle excitations,  for any state based on the abelian hierarchy. We relate our construction to structures in rational conformal field theory and stress the importance of using coherent state wave functions which allows us to formulate an extension of the  bulk - edge correspondence that was conjectured by Moore and Read. Finally, we study the proposed ground state wave functions in the limiting geometry of a thin torus and argue that they coincide with the known exact results.

\end{abstract}

%\submitto{\NJP}
%\maketitle

\section{Introduction}

Most of the quantum Hall states in the lowest Landau level can be understood as part of a hierarchy of states supporting abelian quasiparticles. While there is a recognized and conceptually satisfactory schematic picture for such states \cite{halphi,haldhi}, and a bulk of earlier work based on this \cite{wenzee,fradkinlopez}, many aspects have remained elusive, or else are taken to be intuitively obvious.

In a recent paper \cite{svh10a} we developed methods to construct explicit trial wave functions for quantum Hall hierarchies, prime examples being the abelian hierarchy prominently observed in the lowest Landau level and the non-abelian hierarchies based on it \cite{BS}. The new idea is to note that when expanded in a basis of coherent states, the wave functions for such hierarchical states can be expressed in terms of conformal blocks in conformal field theories. The basic difference from earlier work, is that the local fields representing electrons and quasiparticles include components of both chiralities. While this is necessary to account for hierarchical condensates of quasiholes and their anti-chiral edge modes, the corresponding conformal blocks are related to holomorphic polynomials in the lowest Landau level wave functions only as their representation in the coherent state basis.

In our previous work \cite{svh10a} we gave the physical background and motivations for the construction and presented several examples. The aim of this paper is trifold. First we put the proposed states in the perspective of earlier applications of CFT to quantum Hall physics, and formulate an extended Moore-Read conjecture. Second, we explicitly show how to form the simplest quasihole condensates, namely those  of ordinary holes in a filled Landau level. For these we obtain precisely the states in the negative Jain series, and we propose that all states in the hierarchy can be obtained in a similar way by consecutive condensations of quasielectrons and quasiholes. Third, we deduce the form of these model states on a thin torus, and argue that they agree with the known exact result. Taken together, these results strongly suggest the validity of the model wave functions proposed in \cite{svh10a}. 
  
\section{Hierarchy wave functions and chiral blocks}

Model wave functions for quantum Hall states and their elementary excitations can often be expressed as conformal blocks of chiral operators. A large class of such wave functions are of the form \cite{BS}
\be{RRwf}
  \Psi_\lambda(z_1,\dots, z_N ; \eta_1, \dots, \eta_M)  &=& \Psi_{Ab} (z_1,\dots, z_N ; \eta_1,\dots, \eta_M) \\
   &\times & \,\langle \psi(z_1) \cdots \psi(z_N) \tau (\eta_1) \cdots \tau (\eta_M) \rangle_{\lambda} \, , \nonumber
\ee
where $\Psi_{Ab}$ is an abelian hierarchical wave function and $\langle \cdots \rangle_\lambda$ is a conformal block containing insertions of the operators $\psi$ and $\tau$ associated with electrons and quasiholes. The label $\lambda$ indicates the overall fusion channel of these operators. The focus of this work is on the abelian hierarchy states, for which the second factor is absent. We stress, however, that a general understanding of the functions $\Psi_{Ab}$ is needed in constructing non-abelian hierarchies.\footnote{
There are, however, other approaches to non-abelian QH states. See for example \cite{cappelli, wenwang}.
}

\subsection{Compactified boson, chiral vertex operators and conformal blocks}

An important class of wave functions among $\Psi_{Ab}$ are the hierarchical states that result from sequential condensations of abelian quasielectrons. These can be written in terms of holomorphic conformal blocks of local chiral operators and support edge modes of only one chirality. We refer to such states as chiral.

These chiral states are naturally classified in terms of chiral integral lattices \cite{frohlich1}, which appear in the conformal field theory of compactified massless bosons. More precisely, this structure is present in the chiral half of the theory defined by the action $S[\bvphi] = \frac{1}{8\pi}\int\rmd^2 x \, k_{ij}\partial_\mu\varphi_i\partial^\mu \varphi_j$, invariant under the global $U(1)$ transformations $\varphi_i \to \varphi_i + a_i$ and supporting the conserved chiral currents $J_i(z)=\partial \varphi_i$ and $\bar J_i(z)=\bar \partial\varphi_i$. The operator content of the model can be sorted according to the corresponding charges. Taking the compactification lattice as $\Gamma = \{\mbf e|\mbf e = \sum_{\a=1}^n n_\a \mbf e_\a,\; n_\a\in \mathbb Z,  \a=1,\ldots,n\}$, where $\mbf e_\a$ specify the field identifications $\bvphi  \equiv \bvphi +2\pi \mbf e_\a$, the chiral affine primary fields  of the model assume the form 
\begin{equation}\label{eq:V}
  V_{\mbf q}(z) = \colon \rm e^{\rm i \mbf q\cdot \bvphi(z)}\colon, \;\;\;{\mbf q}\in \Gamma^*
\end{equation}
where $\Gamma^*$ is the dual lattice of $\Gamma$. Hence $V_{\mbf q}(z) $ are periodic on $\Gamma$. The vector $\mbf q$ specifies the $U(1)$ charges and the conformal spin $h_{\mbf q} = \mbf q\cdot \mbf q/2$ of the operator $V_{\mbf q}$ as is implied by the operator product expansions
\begin{eqnarray}
  \label{eq:JV}
  J_i(z)V_{\mbf q}(w) &=& \frac{q_i}{z-w} V_{\mbf q}(w) +\cdots\\
  \label{eq:TV}
  T(z)V_{\mbf q}(w) &=& \frac{\mbf q\cdot \mbf q /2}{(z-w)^2} V_{\mbf q}(w) +
  \frac{1}{z-w}\partial_w V_{\mbf q}(w)+\cdots\\
  \label{eq:VV}
  V_{\mbf q'}(z)V_{\mbf q}(w) &=& (z-w)^{\mbf q\cdot \mbf q'} V_{(\mbf q+\mbf q')}(w)+\ldots,
\end{eqnarray}
where the inner product is defined by $\mbf a\cdot \mbf b = k_{ij}^{-1}a_ib_i$ and $\cdots$ denotes regular terms. In \eref{eq:TV} the stress-tensor $T(z)$  is taken as $T(z) = -\frac{k_{ij}}{2}:\partial \varphi_i \partial\varphi_j:$.

The chiral QH states are those for which the metric on $\Gamma$, given by the scalar product $\mbf e_\a \cdot \mbf e_\b \equiv K_{\a\b}$, is positive-definite and integer-valued with odd or even diagonal elements depending on whether the state is fermionic or bosonic. Formally, the operators $V_{\mbf e_\a}$ generate a lattice \emph{vertex (super)algebra}, whose inequivalent irreducible representations, and hence the primary fields which are of the form \eref{eq:V}, are labeled by charges $\mbf q\in \Gamma^*/\Gamma$ \cite{dong93}.\footnote{
  The story is somewhat more complicated in the case of fermions. However, for QH states the relevant representations are still those given by $\mbf q \in \Gamma^*/\Gamma$.
}
Since the metric $K_{\a\b}$ is integral, the lattices $\Gamma$ and $\Gamma^*$ are commensurate with $\Gamma\subseteq \Gamma^*$, and hence the number of primary fields is \emph{finite} (in fact, it is $d=|\Gamma^*/\Gamma|=\det(K)$). The theory of compactified bosons, so defined, is a rational conformal field theory, whose operator content is completely characterized by the matrix $K$, which also is crucial in Wen and Zee's classification of abelian QH states \cite{wenzee}.

As generators in the vertex algebra, the operators $V_{\mbf e_\a}$ can be interpreted as descendants of the identity operator, a point of view emphasized, for example, in \cite{simon_review}. This formalizes the important notion that the "condensed" particles related to $V_{\mbf e_\a}$ are topologically trivial, \ie they cannot be detected by any holonomy of the primary fields in the model. The basis vectors $\mbf e_\alpha$ in $\Gamma$ are now  identified with the emergent $U(1)$ charges of the condensed particles,\footnote{ 
The existence of these emergent symmetries has been challenged \cite{fradkinlopez}.
}
and the charges $\mbf q$ of the primaries as those of the associated quasihole excitations. The vertex operators carrying the dual charges $\mbf e_\a^*\in \Gamma^*$, which satisfy $\mbf e_\a^* \cdot \mbf e_\b = \delta_{\a\b}$, represent the elementary quasiholes. The LLL QH wave functions are constructed from the chiral blocks of the operators $ V_{\mbf e_\a}(z)$ and $V_{\mbf e_\a^*}(\eta)$, which owing to \eref{eq:VV} are holomorphic in the electron coordinates $z_i$. To connect these formal considerations to the concrete construction given in \cite{hhv08, hhv09}, we recall that at each level in the hierarchy, the quasielectron condensation amounts to introducing a new electron operator that has trivial holonomies with all the quasiparticles, or in physical terms, quasiparticles pick up only trivial phase factors when encircling electrons. 

\subsection{The generalized Moore-Read conjecture and the abelian hierarchy }\label{sec:MR}
Only in the case of the chiral QH states can the relation of model wave functions to integral lattices be established in the straightforward way outlined above. Many prominent states, such as the Jain state at $\nu=2/3$ \cite{wu93}, are naturally described by $K$-matrices that are not positive-definite and therefore have no direct meaning within chiral models. In the edge theories, the negative eigenvalues of $K$ correspond to anti-chiral modes, and in the bulk they signify condensation of quasiholes. A natural ansatz for the bulk wave functions of these states would involve products of holomorphic and antiholomorphic conformal blocks, except for these not being proper LLL wave functions. As  argued and exemplified in \cite{svh10a}, this difficulty is resolved by the following conjecture which generalizes that of Moore and Read: 
\begin{quote}
  {\it Wave functions for QH states, in the basis of coherent states $\ket{\xi_1,\dots ,\xi_N} $, can be expressed in terms of (sums of) products of chiral and antichiral conformal blocks. The corresponding edge modes are given by the same CFTs that describe the bulk wave functions.} 
\end{quote}
By same CFTs, we mean CFTs with the same representation content. In particular, the bulk CFT is fundamentally unable to differentiate between non-propagating and propagating edge modes, as are the corresponding effective topological field theories.

For the particular case of the abelian hierarchy states we have a concrete construction that proceeds as follows. We interpret the $K$-matrix as a Lorentzian lattice metric and decompose it into positive semi-definite, but in general rational-valued, components $\kappa$ and $\bar\kappa$ so that 
\be{decom}
  K = \kappa -\bar \kappa,
\ee
and associate the components with the chiral and antichiral sectors respectively. In this way the monodromies of the generators of the extended algebra are still encoded in $K$. The decomposition is formal in that while the components $\kappa$ and $\bar\kappa$ define independent compactification lattices, it is still $K$ that determines how the representations in these sectors combine to form physical states.

The resulting chiral blocks $\Psi_\kappa$ and $\Psi_{\bar \kappa}$ are not, in general, (anti)holomorphic because of branch cuts of the type $(z_i - z_j)^{\kappa_{\alpha\beta}}$, but still constitute well-defined representations of LLL wave functions. Since $K$ is integral these branch cuts will cancel and thus the expression
\be{spacewf}\fl
  \Psi(z_1,\dots, z_N ) = \int [ d^2\xi_i ]   \bracket {z_1,\dots, z_N}  {\xi_1,\dots, \xi_N}  
 \Psi_\kappa(\xi_1, \dots, \xi_N ) \bar\Psi_{\bar\kappa}(\bar\xi_1, \dots, \bar\xi_N ) 
\ee
is a well-defined holomorphic LLL wave function. As explained in detail in \cite{svh10a} the coherent state kernels $\bracket z \xi$ always combine with the gaussian factors in the blocks  $\Psi_\kappa$ and $\bar\Psi_{\bar \kappa}$ to give, up to a trivial rescaling,  a holomorphic delta function $\delta (z,\xi)$. If the decomposition \eref{decom} is such that both $\kappa$ and $\bar \kappa$ are integer-valued, and hence the chiral blocks $\Psi_\kappa$ and $\Psi_{\bar\kappa}$ (anti)holomorphic, all integrals can be easily evaluated to yield explicit expressions for the wave functions. Also, since $K_{\a\b}$ is precisely the $K$-matrix appearing in Wen's classification of abelian QH states, the (minimal) number of right and left moving edge states is given by the number of positive and negative eigenvalues of $K$ \cite{wencft}. Since the holomorphic and antiholomorphic blocks correspond to the CFTs defined by $\kappa$ and $\bar\kappa$, the chiral and antichiral  edge modes are directly related to the various quasielectron and quasihole condensates respectively.

The decomposition \pref{decom} is not unique as is most easily seen from the example of the $\nu = 1/m$ chiral states, where the one-dimensional $K$-matrix can be decomposed as $m = (m+k) - k$. Taking $k=0$ gives Laughlin's wave function, but any $k$ gives an acceptable wave function with the same topological properties \cite{girvinjach}. It would be interesting to explore if the freedom in the decomposition \pref{decom} could be used to systematically construct improved wave functions for realistic potentials.

\subsection{Quasiparticle condensation and hierarchical electron operators}\label{sec:basis}

We now give explicit expressions for the electron and quasihole operators used to construct a general hierarchical state. For the chiral states this problem has already been solved \cite{hansson07,hhv08,hhv09}, but we shall restate the solution in a form that easily generalizes to the full hierarchy. There is a large freedom in the choice of basis on the charge lattice, reflected in the metric $k_{ij}$. In the previous work this was taken to be diagonal and of the same dimensionality as $K$.  Another choice is $k_{ij} = K_{ij}$, which, implies that the basis vectors of $\Gamma$ and $\Gamma^*$ take the simple form $(\mbf e_{\alpha})_i = K_{\alpha i}$ and   $ (\mbf  e^*_{\alpha})_i = \delta_{\alpha,i}$. Note that since $K_{ij}$ is non-singular the two-point function $\av{ \varphi_i(z)\varphi_j(w) } = -k^{-1}_{ij} \ln (z -w) $ is well defined. In this basis  the electron and quasihole vertex operators  at level $n$ in the hierarchy become
\be{altops}
  V_\alpha &=& \partial_z^{\alpha -1} \rme^{\rmi \sum_\b K_{\a\b}\varphi_\b},\\
  H_\alpha &=& \rme^{\rmi \varphi_{\a}} \, , 
\ee
where $\alpha = 1,...,n$. The derivatives encode the orbital spins of the electrons and thus determine the shift on the sphere \cite{wen} and contribute to the Hall viscosity \cite{read08}. The operators \eref{altops} are not primary with respect to the Virasoro or the extended chiral algebra, but they are descendants of such primary fields. The operators describing the quasielectrons are more complicated quasilocal objects described in \cite{hhv09}, where it is also shown how the derivatives in \pref{altops} appear when these are made to condense. The  discussion of the hierarchical quasiparticle condensation is deferred to \sref{sec:condensation}.

A generalization  to include quasihole condensates, \ie to indefinite $K$-matrices decomposed according to \pref{decom}, is as follows. If both $\kappa$ and $\bar\kappa$ have the same rank as $K$, and hence are invertible, we have
\begin{eqnarray}
  V_\alpha &=& \partial_z^{\sigma_\alpha}\partial_{\zb}^{\bar \sigma_\alpha}  \label{genop}
  \rme^{\rmi \sum_\b\kappa_{\a\b}\varphi_\b}
  \rme^{\rmi \sum_\b \bar \kappa_{\a\b}\bar\varphi_\b},\\
  H_\alpha &=& \rme^{\rmi \varphi_{\a}},\ \ \ \ \ \ \ \ \ \  \label{genhole}
  P_\alpha =  \rme^{\rmi \bar\varphi_{\a}}.
\end{eqnarray}
In this case both quasielectrons and quasiholes can be described by local operators, but this is in general not true if either $\kappa$ or $\bar\kappa$ is not invertible. The above construction can be easily modified, however - an example is given in \cite{svh10a}.

Ground state wave functions in the coherent state basis are now obtained as antisymmetrized conformal blocks of products of the electron operators \pref{genop}. A direct calculation yields 
\begin{equation}\label{eq:planar}
  \Psi_{K,\kappa,s,\sigma} (\{\xi_i\},\{\xib_i\}) = {\cal A} \prod_\a \partial_{\xi_\a}^{\sigma_{\a}} \partial_{\bar\xi_\a}^{\bar \sigma_{\a}}  \prod_{\a<\b}
  (\xi_\a-\xi_\b)^{\kappa_{\a\b}}
  (\xib_\a-\xib_\b)^{\bar\kappa_{\a\b}}      \, , 
\end{equation}
where $\cal A$ is an antisymmetrizer. Here and in the following, gaussian factors are suppressed. The input topological data is the $K$-matrix $K=\kappa-\bar\kappa$ and the spinvector $s = \mbf \sigma - \bar\mbf \sigma$, which specifies the orbital spins of the condensate particles, but the wave function also depends explicitly on the chiral decomposition of $K$ and $s$. We use the short hand notation $(\xi_\a-\xi_\b) = \prod_{i\in I_\a}^{N_\a}\prod_{j\in I_\b}^{N_\b}(\xi_i-\xi_j)$, where $N_\a = n_\a N$ is the number of particles of type $\a$ and $I_\a$ is their index set. The partial fillings $n_\a$ are determined by the homogeneity condition
\be{homogen}
n_\alpha =\frac 1 \nu K^{-1}_{\alpha\beta}\,  t_\b = \frac 1 {\nu_\kappa} \kappa^{-1}_{\alpha\beta}\,  t_\b = \frac 1 {\nu_{\bar\kappa}}\bar\kappa^{-1}_{\alpha\beta}\,  t_\b,
\ee
where $t_\alpha = 1$ and $\nu = t_\a K^{-1}_{\a\b}t_\b$ {\em etc}. Note that this constrains the possible decompositions of the $K$-matrix. In all cases we have considered there are decompositions that solve these constraints, but we have not tried to prove this for an arbitrary hierarchical $K$-matrix. Quasihole and quasielectron states are obtained analogously to \pref{eq:planar} from  blocks containing insertions of the operators \pref{genhole}.

\subsection{Charged and neutral fields}
It is sometimes advantageous to use a formulation where the physical electromagnetic current is singled out among the $U(1)$ currents in the model \cite{read08}. Consider first a system defined by a rank-$d$ positive-definite $K$-matrix $K_{\a\b} = \mbf e_\a \cdot \mbf e_\b$. The electron number current 
\begin{equation}\label{eq:ecurrent}
  J(z) = \frac{1}{2\pi\rmi}\sum_{\alpha} \nu_\alpha \mbf e_{\alpha}\cdot \partial \bvphi(z) \, ,
\end{equation}
where $\nu_\a = \nu \, n_\a$, is such that for all the electron operators $V_{\mbf e_\a}$ the corresponding charge is one. In other words, $\mbf Q\cdot \mbf e_\a = 1$ for all $\a$, where $\mbf Q =\sum_\a\nu_\a \mbf e_\a$ is the charged direction on $\Gamma$ as determined by \pref{eq:ecurrent}. Next we decompose $\mbf e_\a = \mbf e_\a^{c}+\mbf e_\a^{top}$ of $\Gamma$ into rank $1$ and $d-1$ components, satisfying
\begin{eqnarray}
  \mbf Q \cdot \mbf e_\a^{c} &=& 1\\
  \mbf Q \cdot \mbf e_\a^{top}&=& 0
\end{eqnarray}
for all $\a=1,\ldots, d$. Since $\mbf Q\cdot \mbf Q = \nu$, we can take $\mbf e_\a^{c} = \nu^{-1}\mbf Q$. Requiring $\mbf e_\a \cdot \mbf e_\b = K_{\a\b}$, it follows that the neutral components satisfy 
\begin{equation}\label{eq:Ktop}
\mbf e^{top}_\a \cdot \mbf e^{top}_\b = K_{\a\b}-\nu^{-1}t_\a t_\b \equiv
K^{top}_{\a\b},
\end{equation}
which defines what we call a topological $K$-matrix. The vector composed of the partial fillings $n_a $ is a zero-mode of $K^{top}$:
\be{zeromode}
K^{top}_{\alpha\beta} \, n_\beta  = (K_{\alpha\beta}- \nu^{-1}t_\a t_\b )  n_\beta = 0,
\ee
which shows that $K^{top}$ is degenerate. In fact, its rank is $d-1$ as follows from the subadditivity of rank, that is for matrices $A$ and $B$ of the same dimension, $ \mathrm{rk}(A+B)  \le  \mathrm{rk}(A)  + \mathrm{rk}(B)  $. The degeneracy of $K^{top}$ implies that there is a linear relation between the charge vectors $\mbf e_{\alpha}^{top}$. The only consistent such relation is the charge-neutrality condition $\sum_\alpha \nu_\alpha \mbf e_{\alpha}^{top}=0$, which implies that no background charge is required for the neutral sector.  (Any other linear relation would imply an independent zero-mode in contradiction with  $\mathrm{rk}(K^{top}) = d-1$.)

In the case of non-chiral states, the factorization of the charged component can be done similarly, with the right-hand side of \eref{eq:Ktop} again defining the topological $K$-matrix for the neutral sectors. For non-chiral states, such $K^{top}$ is non-chiral and subject to the decomposition \eref{decom}.

\section{Condensation of quasiholes and the abelian hierarchy }\label{sec:condensation}

The condensation of abelian quasiparticles can be implemented in three related ways. 

\subsection{Wave functions in the Haldane-Halperin hierarchy}
In the Haldane-Halperin picture, the wave function for the electronic state containing condensed quasiparticles is written as
\begin{equation}\label{eq:HHwf}
  \Psi^{HH}(\{z_i\}) =\int [\rmd^2\eta_k] %\prod_{i=1}^M\rmd^2\eta_i 
  \Psi_{\mathrm{qp}}^*(\vec \eta_1,\ldots,\vec \eta_M) \Psi(\vec \eta_1,\ldots,\vec \eta_M; z_1,\ldots, z_N),
\end{equation}
where $\Psi$ is the wave function of the parent state of $N$ electrons at a given filling $\nu$ with $M$ quasiparticles at positions $\vec\eta_i$, and $\Psi_{\mathrm{qp}}$ is a suitably chosen pseudo wave function for the quasiparticles. Also, $[\rmd^2\eta_k] = \prod_{k=1}^M \rmd^2\eta_k$.

Let us now consider hole condensates over the filled Landau level. These states are arguably the simplest ones involving quasiparticle condensates since in the LLL approximation they are the exact particle-hole conjugates of Laughlin states. They are obtained by inserting in \eref{eq:HHwf} the wave function \cite{girvin84}
\begin{eqnarray}
  \Psi(\{\eta_k\};\{z_i\}) 
  &=& \rme^{-\frac{1}{4}\sum_{i=1}^M |\eta_i|^2}\prod_{k<l}^M(\eta_k-\eta_l)  \prod_{i,k} (z_i-\eta_k) \prod_{i<j}^N (z_i-z_j),\label{eq:qh1}
\end{eqnarray}
of the $\nu=1$ state of $N$ electrons and $M$ quasiholes, together with an appropriate pseudo wave function. For example, at $\nu=2/3$ the choice
\begin{equation}\label{eq:pseudo}
  \Psi^*_{\mathrm{qp},1/3}(\{\bar\eta_i\}) = \prod_{i<j}^M (\bar \eta_i-\bar \eta_j)^3
  \rme^{-\frac{1}{4}\sum_{i=1}^M |\eta_i|^2},
\end{equation}
with $M=N/2$, gives the hierarchy state that is the exact particle-hole conjugate of the Laughlin state at $\nu=1/3$. Unfortunately, the expression \eref{eq:HHwf} does not simplify and the wave function remains implicit. However, as noted in \cite{mr} it can be related to CFT data, as is easily seen by writing it as
\be{HH23}
  \Psi^{HH}_{2/3}(\{z_i\}) 
  &=&\int [\rmd^2\eta_k]    
  \langle \prod_{k=1}^M \mathcal H(\eta_k,\bar \eta_k) \prod_{i=1}^N
  V(z_i)\rangle,\label{eq:HH23}
\ee
where $\psi(z) = \colon \rme^{\rmi \varphi(z)}\colon$ and 
\begin{equation}\label{eq:Heta}
  \mathcal H(\eta,\bar \eta) = \colon \rme^{\rmi \varphi(\eta) -\rmi \sqrt{3} \bar \phi(\bar\eta)}\colon\, ,
\end{equation}
which can be viewed as a \emph{bosonized} quasihole operator. We define the correlators to include homogeneous background charges  $e^{-i \int d^2 z\,  [\rho_\a\varphi_\a (z) + \bar\rho_\a\bar\phi_a (\bar z)]}$, which make the correlators neutral. Notably, the quasihole operator \eref{eq:Heta} is not purely chiral, consistent with the edge theory of this state. We note also that the normalization factors in \eref{eq:qh1} and \eref{eq:pseudo}, as implied by the construction of the $\nu=2/3$ as the exact particle-hole conjugate of the Laughlin state at $\nu=1/3$, are naturally reproduced by the conformal blocks in \pref{HH23}. Consequently,  the measure $[\rmd^2\eta_k]$ is automatically correct once both $\Psi$ and $\Psi_{{qp}}$ are taken as conformal blocks, a fact that is believed to generalize \cite{mr}.

While \pref{HH23} is a valid ansatz wave function, it suffers from two problems. It obscures the connection to the topological classification of hierarchical states and also has the technical disadvantage of having a non-algebraic form.

\subsection{Hierachy wave functions in coherent state basis}

Given the problems with the conventional hierarchical wave functions, other approaches are needed. As we now show, the generalized Moore-Read conjecture suggested in \sref{sec:MR}, combined with the machinery of conformal field theory, can be used to generate explicit model wave functions for abelian hierarchy states which result from quasihole condensation. This approach generalizes the methods successfully applied to quasielectron condensate states \cite{hhv08}, and generates the full QH hierarchy in the form described in \sref{sec:basis}.

An essential element in our approach is the abstraction of those key elements in the construction of the quasielectron states that enable the efficient use of CFT methods. Crucial there is the use of the charge-conjugate quasihole primary field $H^*$, with its operator-product expansion  $H^*(z)V(w) \sim (z-w)^{-1} (H^* V)(w)$, in the construction of a quasilocal quasielectron operator. In order to construct an analogous quasilocal quasihole operator, one seemingly needs an operator algebra with the OPE
\begin{equation}\label{eq:OPE1}
  P^*(\zb)V(w,\wb) \sim (\zb-\wb)^{-1} (P^* V)(w,\wb),
\end{equation}
where $P$ is the coherent state representation of the charge-conjugate quasielectron. It is clear that in order to realize \eref{eq:OPE1}, the electron operator $V$ has to contain fields of both chiralities. In fact, a filled Landau level allows such a description.

Before proceeding further, let us note that in the case of quasielectron condensate wave functions this seemingly complicated approach is necessary since there is no local quasielectron operator in the CFT spectrum, in contrast to quasihole operators for which a local representation is readily available. This asymmetry reflects the incompressibility of the QH liquids. Quasielectrons, which are local contractions of the liquid, are gapped by correlations in the bulk, while the spectrum of the quasiholes is that of the low-energy edge theory. 
The construction of quasilocal quasielectron operators in \cite{hhv09} allowed one to obtain hierarchy wave functions for which the integrals in \eref{eq:HHwf} become tractable, thereby making possible a direct comparison to composite fermion wave functions \cite{hansson07}. That such an approach also is useful in the description of quasiholes will be seen in what follows. 

To construct a quasilocal quasihole operator, we start with a suitable formulation of the filled Landau level. We write the electron and quasielectron operators, in the basis described in \sref{sec:basis}, as 
\begin{eqnarray}\label{v11}
V_1(\xi, \bar\xi) &=& \colon
\rme^{\rmi \bar\phi_1(\bar\xi) + \rmi 2 \varphi_1(\xi)}\colon,\\
P_1(\bar\eta) &=&   \colon \rme^{\rmi\bar\phi_1(\bar\eta)}\colon.\label{eq:qe}
\end{eqnarray}
The $N$-point block of operators $V_1$ is given by
\begin{equation}\label{eq:psi1}
  \Psi(\{\xi_i\},\{\xib_i\}) = \langle{\prod_{i=1}^N}V_1(\xi_i,\bar \xi_i)\rangle = 
  \prod_{i<j}^N (\xi_i-\xi_j) |\xi_i-\xi_j|^{2} \rme^{-\frac{1}{2}\sum_{i=1}^N |\xi_i|^2}
\end{equation}
which is nothing but a  coherent state representation of the wave function for the filled Landau level \cite{svh10a}. Equivalently, the electron operator can be defined as the chiral quasilocal operator
\begin{equation}\label{eq:psiql}
  {\cal V}_1(z) = \int \rmd^2\xi\, \langle z|\xi\rangle V_1(\xi,\xib)
\end{equation}
where the convolution with the coherent state kernel $\langle z|\xi\rangle = \rme^{-\frac 1 4 (|z|^2-2 \bar \xi z+|\xi|^2)}\equiv \delta_{LLL}(z,\bar \xi)$ projects to the lowest Landau level. This definition reproduces, up to a scale, the wave function of the filled Landau level as the correlator
\begin{equation}\label{eq:psi2}
  \langle \prod_{i=1}^N {\cal V}_1(z_i) \rangle =  \prod_{i<j}^N (z_i-z_j)\rme^{-\frac 1 4 \sum_i |z_i|^2}.
\end{equation}

The operator \eref{eq:qe} is the local quasielectron counterpart to the usual local quasihole operator. In a $\nu=1/m$ Laughlin state ($m>1$), formulated in analogy with the filled LL in eqs. \pref{eq:psi1} and \pref{eq:psi2} above, the insertion of a single quasielectron operator gives 
\begin{eqnarray}
  \Psi(\bar\eta,\{z_i\})&=&\langle P_1(\bar\eta) \prod_{i=1}^N \psi(z_i)\rangle\nonumber\\
  &=& \rme^{-\frac{1}{4}|\eta|^2}\rme^{-\frac 1 4 \sum_{i=1}^N |z_i|^2}\prod_{i=1}^N(2\partial_i - \bar\eta) \, \prod_{i<j}^N  (\bar z_i - \bar z_j)  (z_i-z_j)^{m+1}  \, . \label{eq:lqp}
\end{eqnarray}
Note that the factor $(2\partial_i - \bar\eta)$ is precisely the one appearing in the quasielectron wave function by Laughlin, and in fact \eref{eq:lqp} differs from it only by a local correlation factor $\prod_{i<j}^N  |z_i-z_j|^{2}$, which does not affect the filling fraction.

To construct quasihole condensates, the idea, then, is to represent the quasiholes by $P^*$ fused with an electron operator in full analogy to the construction of the quasielectron operator \cite{hhv09}. More precisely, we define a \emph{quasilocal} (quasi)hole operator
\begin{equation}\label{qlqh}
  {\cal H}(\eta) = \int d^2 w\,\langle \eta |w\rangle
 \colon P^*(\bar w)\partial \bar J(\wb) \colon .
\end{equation}
The operator \eref{qlqh} can be thought of as creating a quasilocal hole as a superposition of expanded correlation holes around electrons in the vicinity of $\eta$. The divergence of the current, $\partial \bar J$, has support only at the electron (and quasihole) positions, as result of which the hole operator $P^*$ fuses with the electron operators when inserted into a correlator. Further details of this approach and the precise meaning of the normal ordering symbol $\colon\ \colon$ are explained in \cite{hhv08} and \cite{hhv09}. This construction may seem unnecessarily complicated, given the success of the simpler Laughlin hole. However, it does put quasiholes and quasielectrons on equal footing, and, as we will show, solves the long-standing problem of an explicit quasihole condensate construction, resulting in (as a special case) precisely the wave functions in the  negative Jain series.

Before addressing condensates, let us first simply consider a single-hole state. In addition to making us more familiar with our new object, this will show that this approach produces {\it quantitatively} good hole states. The wave function for the single-hole state can be written as
\be{oneqecorr}
\Psi(\eta; \{ z_i \}) &=& \langle{\cal H}(\eta)\prod_i \mathcal{V}_1(z_i)\rangle\\
&=&\int [\rmd^2\xi_i]\langle z_i|\xi_i\rangle \int\rmd^2w\langle \eta|w\rangle
\langle\colon P^*(\wb)\partial \bar J(\wb)\colon  \prod_i V_1(\xi_i,\bar\xi_i) \rangle \, .\label{eq:psi1h}
\ee
The correlator inside the integral can be evaluated using the definition of the normal ordering and the $U(1)$ Ward identity, with the result 
\begin{equation}\fl
  \langle\colon P^*(\wb)\partial \bar J(\wb)\colon  \prod_i V_1(\xi_i,\bar\xi_i) \rangle= \sum_{k=1}^N (-1)^k\delta^2(w-\xi_k)
  \langle \colon P^*(\xib_k) V_1(\xi_k,\xib_k) \colon \prod_{i\neq j} \, 
  V_1(\xi_j,\xib_j)\rangle\label{eq:tJpsi}.
\end{equation}
Using the methods described in \cite{hhv09}, the  normal-ordered product is evaluated as
\be{currvert}
  \colon P^*(\xib) V_1 (\xi,\xib)\colon  = \bar \partial  \colon\rme^{\rmi 2\varphi_1(\xi)}\colon
\ee
Note that because of the presence of the zero mode, and the associated need for compensating charges to define neutral correlators, the chiral vertex operators are not strictly  holomorphic. Using constant background charges, this is manifested in the gaussian factors appearing in the correlators. More generally, when a coherent state wave function $ \Psi(\{\xi_i\},\{\xib_i\}) $ in \pref{eq:planar} is turned into a real space LLL wave function $\Psi(\{z_i\})$ by convoluting it with the kernels $\langle z_i|\xi_i\rangle$, the derivatives are most easily evaluated by partial integrations, which effectively enforces the substitutions $\partial_\xi \rightarrow \partial_z/2$ and $\partial_{\bar\xi} \rightarrow -z/4$. This procedure avoids the introduction of the covariant derivatives which are necessary when using the conformal blocks to directly give the real space wave functions \cite{hhv09}.

By combining the result of \eref{eq:tJpsi} with \eref{eq:psi1h} and  using that the coherent state projection identifies $\bar\xi = 2\partial$, we get
\begin{equation}
\Psi(\eta; \{ z_i \}) =\rme^{-\frac 1 4 |\eta|^2}
\rme^{-\frac 1 4 \sum_{i=1}^N |z_i|^2}
\sum_{k=1}^N (-1)^k \rme^{\eta \partial_k} z_k \prod_{i<j}^{(k)}(\partial_i - \partial_j)\prod_{i<j}(z_i - z_j)^2\label{1hb} .
\end{equation}
where the factor $z_k$ originates from a contraction with  the background charge operator, and the superscript $(k)$ indicates that coordinate $k$ is omitted in the product. Setting the hole position $\eta$ equal to zero, one finds that \eref{1hb} is {\it identical} to the "composite fermion antiparticle", which was studied numerically for up to 10 particles in \cite{jeon05} and was found to be closer to the Coulomb hole state, in terms of energies, overlaps and charge profiles, than Laughlin's quasihole. Similar conclusions apply for the corresponding quasihole in the $\nu = 1/3$ Laughlin state. It thus appears that our quasilocal quasihole operator \eref{qlqh} provides a qualitatively and quantitatively reasonable starting point for a hole condensate, which we now turn to. 

Using \eref{qlqh} as a starting point, we can write a hierarchy wave function analogous to \eref{eq:HH23}
\begin{equation}\label{hier23}
  \Psi_{2/3}(\{z_i\})  =  %{\cal A}\left[ \{\bar\xi \}, \{ z \} \right]  
  \int [d^2 \eta_k]\, \langle \prod_{k=1}^{N/2} {\cal H} (\eta_k) \prod_{i=1}^N   \mathcal{V}_1(z_i) \rangle \, ,
\end{equation}
where $\mathcal H$ is a bosonized version of \eref{qlqh}, such that the auxiliary field generates an appropriate pseudo wave function for the quasiholes. The evaluation of the correlator proceeds similarly to the above case of a single quasihole insertion. One finds that the $\eta$-dependent factors combine to (anti)holomorphic delta functions $\exp[-\sum_j(|\eta_j|^2 - \bar\xi_j \eta) ]$, making the $\eta$ integrals trivial. Hence, in this formulation, the unmanageable $\eta$ integrals in \eref{eq:HH23} have been traded for the LLL projections implicit in the quasilocal operators $\mathcal H(\eta_k)$ and $\mathcal{V}_1(z_i)$. A benefit of this approach is that the remaining integrals can also be done, giving exactly the composite fermion wave function for $\nu=2/3$ \cite{wu93}. Additionally, in the coherent state basis, the wave function can be expressed as an antisymmetrized correlator of local operators $V_\a$, in which the topological data, $K = \kappa-\bar\kappa$, and the orbital spins are explicit. For the $\nu=2/3$ wave function, we therefore get the following coherent state wave function,
\begin{equation}\label{hier23b}
  \Psi_{2/3}(\{\xi_i,\bar\xi_i\}) =  %{\cal A}\left[ \{\bar\xi \}, \{ z \} \right] 
{\cal A}  \langle  \prod_{i\in I_1}^{N/2}   V_{1}(\xi_i,\bar\xi_i)   \prod_{j\in I_2}^{N/2}   V_{2}(\xi_i,\bar\xi_i) \rangle  \, ,   
\end{equation}
where $V_2=\bar\partial e^{i \bar\phi_2+\rmi 2 \varphi_1}$ is the electron operator at level 2 of the hierarchy. This construction naturally generalizes to other quasihole condensates, and combined with the quasielectron codensation described in \cite{hhv09}, it generates all abelian hierarchy states as correlators of the operators \pref{genop} given in Section 2.3.

\subsection{Algebraic formulation}

The approach described above demonstrates that hierarchy wave functions, for non-chiral states, can be expressed in terms of conformal blocks of local operators. These operators are generated hierarchically as the fusion products of the suitably defined condensing quasiparticle and the electron operator. As topologically trivial condensate operators, they enable a direct connection to the classification of the abelian quantum Hall states in terms of chiral integral lattices.

In order to elucidate the general structure, it is again convenient to use the basis given in \sref{sec:basis}. Let us first consider a transition between two chiral states, for example the formation of the $\nu=2/5$ state as a condensate of quasielectrons in the Laughlin state at $\nu=1/3$. At $\nu=1/3$ the local electron and quasihole operators are given by 
\begin{eqnarray}
  V_1 &=& \colon \rme^{\rmi 3\varphi_1}\colon\\
  H_1 &=& \colon \rme^{\rmi \varphi_1}\colon,
\end{eqnarray}
which imply the following local quasielectron operator
\begin{equation}\label{eq:P1}
  P_1 = \colon H_1^* V_1\colon =\partial \colon \rme^{\rmi 2 \varphi_1}\colon.
\end{equation}
Upon condensation the quasielectron degree of freedom becomes topologically trivial, which is reflected algebraically in that the corresponding creation operator extends the chiral symmetry algebra. Since only local (\ie with trivial monodromies) operators can condense, an auxiliary field must be introduced to adjust the self-monodromy (spin) of $P_1$, without changing the relations of the underlying vertex algebra --- in effect to \emph{fermionize} $P_1$. The condensing operator, then, is given by
\begin{equation}
  V_2  = P_1 \colon\rme^{\rmi x \varphi_2}\colon 
\end{equation}
where $\varphi_2$ is the auxiliary field. This auxiliary field introduces a new chiral current in the system and increases the dimension of the charge lattice, thereby transforming the $K$-matrix as
\begin{equation}
  K_{\nu =1/3}= (3)\to  K_{\nu = \frac{x-1}{3x-4}} =\left(\begin{array}{cc} 3&2\\2&x
  \end{array}\right),
\end{equation}
where $x$ determines the condensate filling fraction. The minimal choice $x=3$ corresponds to a $\nu=1$ condensate of $e/3$ charged quasiparticles and leads to the hierarchy state at $\nu=2/5$, while \eg the choice $x=5$ gives a state at $\nu=4/11$.

The simplest example of a quasihole condensation is the transition from a filled Landau level at $\nu=1$ to the hierarchy state at $\nu=2k/(2k+1)$ by a condensate of holes. In the coherent state basis, the local electron and quasiparticle operators are 
\begin{eqnarray}
  V_1 &=& \colon \rme^{\rmi 2\varphi_1+\rmi \bar \phi_1}\colon\\
  P_1 &=& \colon \rme^{\rmi \bar\phi_1}\colon.
\end{eqnarray}
Just as in \pref{currvert}  we derive the local quasihole operator  
\begin{equation}\label{eq:H1}
  H_1= \colon P_1^* V_1\colon =\bar \partial  \colon \rme^{\rmi 2\varphi_1}\colon .
\end{equation}
Again, introducing a fermionizing auxiliary field, the new condensate operator can be written as $V_2 = H_1 \colon \rme^{\rmi x \bar \phi_2} \colon$. The $K$ matrix for this extended theory then becomes
\begin{equation}
  K_{\nu =1}= (1)\to  K_{\nu = \frac{x+1}{x+2}} =\left(\begin{array}{cc} 1&2\\2&2-x
  \end{array}\right).
\end{equation}
so $x=2k-1$ gives the sequence $\nu = 2/3, 4/5, 6/7 \dots$. The hierarchy of $K$-matrices, and the corresponding condensate vertex operators, can be generated by iterating this scheme.

We conclude by noting the quasiparticle condensation as formulated here can be regarded as an example of the more general theory of topological phase transitions formulated in \cite{bais}. There is a subtlety, however, since the particle content of the hierarchy states typically does not include monodromy-free operators, apart from those corresponding to the condensates, and hence the condensation can only be effected by introducing an auxiliary field.

\section{Wave functions on the thin cylinder}

A useful test of model wave functions for quantum Hall states is to compare them with the results obtained in the  Tao-Thouless (TT) limit, \ie on a manifold with the topology of a torus or a cylinder with a small compactification radius and a flat metric. In this limit the quantum Hall problem is exactly solvable and the ground state is determined by few generic properties of the interaction potential \cite{bekar}. This universality makes it meaningful to compare ansatz wave functions with the corresponding exact solutions. 

A straightforward way to obtain the cylinder analogue of the general wave function \pref{eq:planar} is to substitute $\xi_\a=\rme^{2\pi\rmi \omega_\a/L}$, where $\omega$ is the complex coherent state parameter on the cylinder \cite{bekar}.\footnote{
  Such a substitution assumes implicitly that the wave functions 
  transform covariantly under conformal transformations. This is only true for the
  coherent state wave functions \eref{eq:planar}, not their real-space images as holomorphic polynomials.
} 
The resulting wave function will depend on the particle numbers $N_\alpha$, which are determined up to order $\mathcal O(N)$ by  $K_{\a\b}$ with order $\mathcal O(1)$ shifts depending on boundary conditions. For chiral states, the periodic patterns emerging in the TT limit turn out to be independent of these shifts, and also of the number of derivatives $ \partial^{s_{\a}} $.  This is not true for the more general states \pref{eq:planar}. Rather than trying to impose boundary conditions on the cylinder functions, we shall derive them as the limit of the torus wave functions \cite{torus}. On a closed manifold, the electron numbers $N_\a$ are fully determined, and on the torus the derivatives turn into finite translations that only provide phase factors \cite{torus} which are of no importance here since the TT-limit is determined by the density profile of the state. 

To find the TT-limit of the cylinder analogue of \eref{eq:planar}, we first consider the known case of a \emph{chiral} state \cite{bergh}. In this case, on a cylinder of circumference $L$ the ground state wave function is given by
\begin{eqnarray}
  \Psi_{\kappa,\mbf s}(\{\omega_i\}) 
  &=&
  \prod_{\a<\b} \sin\left(\frac{\pi}{L}(\omega_\a-\omega_b)\right)^{\kappa_{\a\b}}
  \\
  &=& 
  \prod_\a \xi_\a^{-N_\Phi/2+\kappa_{\a\a}/2}\prod_{\a<\b}(\xi_\a-\xi_\b)^{\kappa_{\a\b}},
  \label{eq:cylinder}
\end{eqnarray}
where the first line is simply  the cylinder holomophic block  of the same {\em primary} operators that were used to obtain  \pref{eq:planar}, and the second line follows from the relation $\sum_\b K_{\a\b} N_\b =N/\nu =   N_\Phi  $ which holds on the torus for all $\a$. Further details on the cylinder-torus connection will be given elsewhere \cite{juhathesis}. It is useful to expand \eref{eq:cylinder} in a Fock basis $\{\ket{\mbf k}\equiv \ket{k_1\cdots k_N}\}$ composed of normalized single-particle states
\begin{equation}\label{eq:sp}
  \psi_k(x,y) = \frac{1}{\pi^{1/4}\sqrt{L}} \rme^{2\pi\rmi k x /L} \rme^{-(y+2\pi k/L)^2/2}
  = \frac{\rme^{-(2\pi k/L)^2/2}}{\pi^{1/4}\sqrt{L}} \xi^k \rme^{-y^2/2},
\end{equation} 
where, again, $\xi= \rme^{2\pi\rmi(x+\rmi y)/L}$ and $k\in \mathbb Z_{N_\Phi}$, with $\mathbb Z_{N_\Phi}= [-N_\Phi/2+\kappa_{\a\a}/2, N_\Phi/2-\kappa_{\a\a}/2]$. In this basis the state \eref{eq:cylinder}, up to an overall normalization, reads
\begin{equation}\label{eq:fock}
  \ket{\Psi} = \sum_{\mbf k} c_{\mbf k}\, \rme^{\half \left(\frac{2\pi {\mbf k}}{L}\right)^2}
  \ket{\mbf k},
\end{equation}
where the sum is over partitions $\mbf k$ of the total momentum $K=0$ of \eref{eq:cylinder} into $N$ unequal (half-)integers in the range $\mathbb Z_{N_\Phi}$, and the coefficients $c_{\mbf k}$ are independent of $L$.

In the TT-limit, $L\to 0$, the expansion \eref{eq:fock} simplifies drastically. It is dominated by the Fock state(s) with the largest $\mbf k^2 = \sum_{i=1}^N k_i^2$ and for which $c_{\mbf k}\neq 0$; other configurations are exponentially suppressed relative to it. On a thin cylinder, the state \eref{eq:cylinder} therefore reduces to an uncorrelated product state, or a small-dimensional linear combination of such states. The eigenstates of a repulsive two-body potential in the lowest Landau level are also of this form. The question is whether the Fock expansion \eref{eq:fock} reduces to the ground state configuration or not.

This question can be answered by simply determining the dominant configuration in \eref{eq:fock} and comparing to the exact ground state. An implicit solution to the former problem is given by
\begin{equation}\label{eq:domi}
  \hat k_i = N_\Phi/2-\min_\alpha \left(h_{\a} +\sum_{j=1}^{i-1} K_{\a\hat{\a}_j}\right),
\end{equation}
where $\hat\alpha_j$ is the type of the particle carrying the momentum $\hat k_j$. Since the polynomial \eref{eq:cylinder} is homogeneous, carrying a fixed total momentum, the single-particle momenta of the dominant configuration $\hat \mbf k$ maximize the variance of $\mbf k$ under the constraint $c_{\mbf k}\neq 0$. This solution is to be compared with the ground state on the thin cylinder \cite{bekar}
\begin{equation}
  k_i^0 = N_\Phi/2-[i/\nu],
\end{equation}
where $i=0,\ldots, N-1$ and $[x]$ is the integer closest to $x$.

For a \emph{non-chiral} state, the problem is harder. A real-space LLL wave function of such a state can be obtained by expanding the wave function as monomials and projecting to the  LLL by convoluting them with the cylinder coherent state kernel. The rule for this is
\begin{equation}
  \xi^{k} \xib^{\bar  k} \to \rme^{\left(\frac{2\pi}{L}\right)^2
    \left(k^2-\frac12(k-\bar k)^2\right)} \psi_{k-\bar k}(z),
\end{equation}
where $\psi_{k-\bar k}$ is the LLL state with momentum $k-\bar k$ as given in \eref{eq:sp}. This implies the following Fock expansion for the antisymmetrized wave function:
\begin{equation}
  \ket{\Psi} = \sum_{\mbf k,\bar \mbf k} c_{\mbf k}
  \bar c_{\bar \mbf k} \rme^{\left(\frac{2\pi}{L}\right)^2
    \left(\mbf k^2-\frac12(\mbf k-\bar \mbf k)^2\right)} \,
  \ket{\mbf k-\bar \mbf k}.
\end{equation}
Again, for $L\to 0$, this expansion is dominated by the Fock state whose momenta $\hat{\mbf k}$, $\hat{\bar \mbf k}$ maximize $f(\mbf k,\bar \mbf k) =  \mbf k^2-\frac12(\mbf k-\bar \mbf k)^2$ and for which the expansion coefficient $c_{\mbf k} \bar c_{\bar \mbf k}\neq 0$. Finding the global maximum of $f$ with this constraint is a non-trivial problem. We argue, however, that the dominant configuration can be construed from the dominant $\hat \mbf k$ and $\hat {\bar \mbf k}$, of the respective chiral components,  as the \emph{ordered} difference $\hat \mbf k-\hat {\bar \mbf k}$.\footnote{For $\mbf k$ and $\bar \mbf k$ ordered as $k_i\geq k_{i+1}$ and $\bar k_i\geq \bar k_{i+1}$, the variation of $f$ upon permutation of two labels, say $\bar k_1 \leftrightarrow \bar k_2$, is $\delta f = -(k_1-k_2)(\bar k_1 -\bar k_2) \leq 0$. This suggests that the momenta $\hat\mbf k$ and $\hat{\bar \mbf k}$ should be paired as ordered lists, although this argument does not exclude the possibility that some higher order permutation gives configuration of higher weight.} In all cases we have checked, the configuration so constructed is the thin-cylinder ground state.

A few examples are in order.
\begin{itemize}
\item The hierarchy state at $\nu = 2/3$, according to the algorithm outlined above, reduces to a Fock state with a periodically repeated unit cell $\ket{1\,1\,0}$, which is, in fact, the unique pattern of that periodicity at this filling and 
agrees with the electrostatic ground state. 
In our algorithm, this TT-pattern is
construed out of the chiral patterns $\ket{101010\cdots}$ and ${\ket{\bar 2\bar 2\bar 2\cdots}}$.  
Notably, the naive ansatz wave function $\Psi = (1-1)(2-2)(1-2)^2$ has a phase-separated TT-limit.

\item The hierarchy state at $\nu = 5/7$, realized as a quasielectron condensate over $\nu=2/3$, is characterized by the $K$-matrix
\begin{equation}
  K =   \left( 
  \begin{array}{ccc} 
    1&2&2\\
    2&1&0\\
    2&0&1
  \end{array}
  \right) 
  =
  \left( 
  \begin{array}{ccc} 
    3&2&2\\
    2&5&2\\
    2&2&5
  \end{array}
  \right)
  -
  \left( \begin{array}{ccc} 
    2&0&0\\
    0&4&2\\
    0&2&4
  \end{array}
  \right).
\end{equation}
In the TT-limit, this state reduces to the configuration $\ket{1\,1\,0\,1\,1\,1\,0}$, which, again, coincides with the electrostatic ground state.

\item Finally, the level-four state at $\nu=8/21$, which arises from the $\nu=2/5$ state by two consecutive quasihole condensations, is  characterized by the $K$-matrix
\begin{equation}
  K =   \left( 
  \begin{array}{cccc} 
    3&2&2&2\\
    2&3&4&4\\
    2&4&3&4\\
    2&4&4&3
  \end{array}
  \right) 
  =
  \left( 
  \begin{array}{cccc} 
    4&2&2&2\\
    2&8&4&4\\
    2&4&6&6\\
    2&4&6&6
  \end{array}
  \right)
  -
  \left( \begin{array}{cccc} 
    1&0&0&0\\
    0&5&0&0\\
    0&0&3&2\\
    0&0&2&3
  \end{array}
  \right).
\end{equation}
This gives as the thin cylinder limit the periodic structure $ABABA$ with $A = \ket{ 0\, 1\,0\,0\,1}$ and $B = \ket{0\,0\,1}$, again agreeing with the electrostatic ground state. In this case, the thin cylinder pattern is sensitive to the chiral decomposition of the $K$-matrix.
\end{itemize}

Given the above examples, it is suggestive that $\hat{\mbf k} -\hat{\bar \mbf k}$ is indeed the dominant configuration. We now give arguments to support this. If we consider only the holomorphic part, all other configurations in the Fock expansion \pref{eq:fock}, can be obtained by "squeezing"\footnote{
An elementary squeeze of $\vec k$ is a change $k_i \to k_i + 1$ and $k_j \to k_j -1$ where $k_i < k_j+1$. A general squeeze is obtained from a sequence of elementary squeezes.
}
the dominant configuration while keeping the ordering of the particles. The variation of the cost function $f$ due to a squeeze $\hat{\mbf k}\to \hat{\mbf k} +\mbf k_\delta$ is 
\be{holsq}
  \delta f = \half \mbf k_\delta^2 +\mbf k_\delta\cdot \hat{\mbf k}+\mbf k_\delta\cdot \hat{\bar \mbf k} \, .
\ee
This variation is negative since the first two terms are a variation of the chiral cost function due to a squeeze and $\mbf k_\delta \cdot \hat{\bar \mbf k} <0$ since $\hat{\bar \mbf k}$ is ordered. The corresponding squeeze in the antichiral sector gives the variation
\be{aholsq}
  \bar \delta f = -\half \bar \mbf k_\delta^2+\bar \mbf k_\delta\cdot (\hat{\mbf k}-\hat{\bar \mbf k}) \, .
\ee
This too is negative provided that the  momenta in  $\mbf k-\bar \mbf k$ are  positively ordered. This is true in all cases we have looked at, but we do not have a proof for a general hierarchical $K$-matrix. The variation due to simultaneous change in $\mbf k$ and $\bar \mbf k$ is $\mbf k_\delta\cdot (2\mbf k + \mbf k_\delta)/2 +\bar \delta f +\mbf k_\delta\cdot (\bar\mbf k +\bar \mbf k_\delta)$, which is always negative, since the first term is the change in $\mbf k^2/2$ and the last is  negative since $\bar\mbf k +\bar \mbf k_\delta$ is ordered. These properties imply that $\hat{\mbf k} -\hat{\bar\mbf k}$ is a \emph{local} maximum of the cost function.

This argument can be applied  iteratively to any state obtained by one, two, etc. squeezes, again provided that the  momenta in  $\mbf k-\bar \mbf k$ remain  positively ordered after every step. To turn these plausibility arguments into a full proof that would require both showing that for any hierarchical $K$-matrix there exist a decomposition such that $\hat{\mbf k} -\hat{\bar \mbf k} $, is positively ordered and that this ordering remains under arbitrary squeezes.

An alternative approach is to maximize the total cost function $f(\mbf k,\bar \mbf k) = \mbf k^2 -\half (\mbf k-\bar \mbf k)^2$ term-by-term, in analogy with the procedure outlined in \cite{bergh}. If at each step the range of possible $k_i$ is larger than that ot $\bar k_i$ then $f(k_i, \bar k_i)$ is maximized by $\max k_i^2$ and $\max \bar k_i^2$, leading to the same result as above (assuming here that we are using shifted momenta that are in a positive range).

Finally we note that the reduction algorithm \eref{eq:domi} giving the dominant monomial can be applied separately for the charged and topological parts defined by \eref{eq:Ktop}, so that the dominant monomial is obtained as an ordered product of dominant monomials of the two constituent parts. This is potentially useful because the dominant monomial of the charged factor is trivial; scaling away the constants, we get  $k_c = (0, \nu^{-1}, 2\nu^{-1},\ldots, (N-1)\nu^{-1})$. It is therefore the neutral part that determines the length of the unit cell. 

\section{Conclusions and outlook}

In this and the accompanying \cite{svh10a} paper we have given a comprehensive description of the full abelian QH hierarchy, that allows us to construct explicit algebraic expressions for a general hierarchical state given its $K$-matrix. An obvious, and important, extension of this work is to construct  the wave functions on the sphere and on the torus. The former will ensure that the shifts, as calculated in \cite{svh10a} are the correct ones, and the latter might open for a direct calculation of the quantum Hall viscosity.

There are also several open questions. For the abelian states, it is not clear what are the most general allowed decompositions $K =\kappa - \bar\kappa$ and  $s = \sigma - \bar\sigma$  that corresponds to good quantum Hall states. Although homogeneity and consistency with the exact results in the TT limit do give restrictions, on $\kappa$ and $\bar\kappa$, and our explicit hierarchy construction suggests definite values for $\sigma$ and $\bar\sigma$, it is quite clear that there are many combinations that give different acceptable representative wave functions with the same topological properties. It would be interesting to investigate to what extent this freedom could be used to find optimized linear combinations for a given potential.

The most intriguing questions, however, concern non-abelian hierarchies, which were briefly touched upon in \cite{svh10a}. In particular it would be interesting to investigate the relationship of our construction to Wen's alternative description of non-abelian states using the parton construction \cite{wenparton}, and also whether the recently proposed mechanism for condensation of  non-abelian quasiparticles \cite{hermanns10} could be used to generate genuinely new, non-abelian hierarchies. 

\ack We would like to than Eddy Ardonne,  Emil Bergholtz and Anders Karlhede for 
discussions and comments on the manuscript.

\end{document}